\documentclass[conference]{IEEEtran}
\usepackage{cite}
\usepackage{graphicx}
\usepackage{tikz}
\usepackage{tikz-3dplot}
\usetikzlibrary{shapes.geometric}
\usetikzlibrary{decorations.pathreplacing}
\usetikzlibrary{positioning}
\usepackage[top=.81in, bottom=.81in, left=.81in, right=.81in]{geometry}  
\usepackage[linesnumbered,algoruled,boxed,lined]{algorithm2e}
\usepackage{algpseudocode}
\usepackage[absolute]{textpos}
\usepackage{threeparttable}
\usepackage[strict]{changepage}
\usepackage{bm,upgreek} 
\usepackage{amssymb}
\usepackage{bbm}
\usepackage{hhline}
\usepackage{amsthm}

\ifCLASSINFOpdf
\else
\fi
\usepackage{amsmath}
\usepackage{amsfonts}
\usepackage{amssymb}

\ifCLASSOPTIONcompsoc
  \usepackage[caption=false,font=normalsize,labelfont=sf,textfont=sf]{subfig}
\else
  \usepackage[caption=false,font=footnotesize]{subfig}
\fi
\usepackage{url}

\begin{document}

%

%
\title{Q-Learning Algorithm for VoLTE Closed Loop Power Control in Indoor Small Cells}

%
%
%


\author{\IEEEauthorblockN{Faris B.~Mismar and Brian L.~Evans}

\IEEEauthorblockA{Wireless Networking and Communications Group, The University of Texas at Austin, Austin, TX 78712 USA}

}

\maketitle

\begin{abstract}
We propose a reinforcement learning (RL) based closed loop power control algorithm for the downlink of the voice over LTE (VoLTE) radio bearer for an indoor environment served by small cells.  The main contributions of our paper are to 1) use RL to solve performance tuning problems in an indoor cellular network for voice bearers and 2) show that our derived lower bound loss in effective signal to interference plus noise ratio due to neighboring cell failure is sufficient for VoLTE power control purposes in practical cellular networks.  In our simulation, the proposed RL-based power control algorithm significantly improves both voice retainability and mean opinion score compared to current industry standards.
The improvement is due to maintaining an effective downlink signal to interference plus noise ratio against adverse network operational issues and faults.
\end{abstract}


%
\IEEEpeerreviewmaketitle

\section{Introduction}

Wireless networks are vulnerable to operational issues, faults and network element failures.  In addition, wireless transmission can face blockage, interference, and other impairments.  While cellular data applications are made resilient against wireless impairments through modulation, coding, and retransmissions, delay-sensitive applications such as voice or low latency data transfer may not always benefit from retransmission since it increases delays and risk of data duplication. Therefore, these applications must be made resilient through other means.

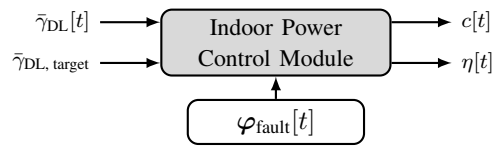
\begin{figure}[!t]
\begin{adjustwidth}{.25in}{0cm}
\centering
\begin{tikzpicture}[style=thick,scale=0.8]
	\node [rectangle, draw, rounded corners, 
		text width=8em, text centered, minimum height=1em, fill=gray!30] at (0,0) (pc) {\small Indoor Power Control Module};
	\node [rectangle, draw, rounded corners, 
		text width=6em, text centered, minimum height=1em, below= of pc,yshift=2em] (fault) {$\bm{\varphi}_\text{fault}[t]$};

	\path [draw, latex-] (pc.170) -- ++(-1,0) node[left] {\smaller $\bar\gamma_\text{DL}[t]$};
	\path [draw, latex-] (pc.190) -- ++(-1,0) node[left] {\smaller $\bar\gamma_\text{DL, target}$};

	\path [draw, -latex] (pc.10) -- ++(1,0) node[text width=5em] [right] {\smaller $c[t]$};
	\path [draw, -latex] (pc.350) -- ++(1,0) node[text width=5em] [right] {\smaller $\eta[t]$};
	\path [draw, -latex] (fault.90) -- (pc.270);
\end{tikzpicture}%
\end{adjustwidth}
\caption{Downlink power control module. $\bar\gamma_\text{DL}[t]$ is the effective signal to noise plus interference ratio (SINR) at the receiver at time $t$ fed back to the power control module.  The module maintains the downlink SINR at the receiver at $\bar{\gamma}_\text{DL, target}$ amid the faults captured in $\bm\varphi_\text{fault}[t]$.  This is done through a power control command $c[t]$ and a repetition factor $\eta[t]$.}
\label{fig:overall}
\end{figure}

The received \textit{signal to interference plus noise ratio} (SINR) is a critical quantity to ensure resilient communications in applications such voice.  In this paper, we propose a framework to automatically tune a cellular network through the employment of reinforcement learning (RL). We devise an RL-based algorithm to improve downlink SINR in an indoor environment for packetized voice using  \textit{power control} (PC) as shown in Fig.~\ref{fig:overall}.  The technology of focus is the fourth generation of wireless communications or \textit{long term evolution} (4G LTE) or fifth generation of wireless communications (5G). 

Downlink closed loop power control is last implemented in 3G \textit{universal mobile telecommunications system} (UMTS) \cite{3gpp25214}.  It rapidly adjusts the transmit power of a radio link of a dedicated traffic channel to match the target DL SINR.  This technique is not present in 4G LTE or 5G due to the absence of dedicated traffic channels for packet data sessions.  However, the introduction of \textit{semi-persistent scheduling} (SPS) in 4G LTE has created a virtual sense of a dedicated downlink traffic channel for VoLTE on which a closed loop power control can be performed.  This scheduling is at least for the length of one voice frame---which is in order of tens of LTE \textit{transmit time intervals} (TTIs).

An improved decentralized $Q$\nobreakdash-learning algorithm was used to reduce interference in the LTE femtocells environment in \cite{5983301}. Various power control algorithms including open loop power control were compared with this $Q$\nobreakdash-learning based approach.
Combining information theory with machine learning, a deep reinforcement learning method  was used in \cite{infoml}.  This method was a constrained optimization problem to maximize the $Q$\nobreakdash-function using the Kullback-Leibler divergence and entropy constraints instead of exploration, which we used in this paper.  Closed loop power control implementation for LTE with the employment of fractional path loss compensation was done in \cite{5638376}.  This resulted in an improved system performance.  However, there was no reference to machine learning or RL in general.   
$Q$\nobreakdash-learning based power control for indoor LTE femtocells with an outdoor macro cell was studied in \cite{7442078}.  The focus was on throughput and the UE reported SINR or call quality indicator was used to change the femtocell transmit power.  To resolve the issue of communicating base stations, a central controller was introduced.  We confine the geographical area to make it is feasible for the base stations to communicate through the backhaul.  Furthermore, an assumption that downlink power control is achieved over shared data channels, which is a relaxed assumption and requires that the scheduler is aware ahead of time about the channel condition for the upcoming user to perform power control, was made.  We did not make this assumption to maintain a more realistic environment since we exploit the use of SPS in packetized voice. 

 While the focus of our paper is on \textit{voice over LTE} (VoLTE),  any future technology offering packetized voice can benefit from our proposed algorithm.  In fact, with the highly anticipated role of \textit{self-organizing networks} (SON) in 5G \cite{6963801}, similar algorithms can be a significant step towards machine learning enabled SON.
 
 Our main contributions are as follows:
\begin{enumerate}
\item Use RL to solve performance tuning problems in an indoor cellular network for voice bearers.
\item Show that our derived lower bound loss in effective SINR due to neighboring cell failure is sufficient for VoLTE power control in practical cellular networks.

\end{enumerate}

\section{System Model}
\label{sec:sysmod}
The system comprises two components:
\begin{enumerate}
\item A radio environment where VoLTE capable UEs are served.
\item A reinforcement learning model using $Q$-learning to perform closed loop power control to improve effective DL SINR measured at the receiver.
\end{enumerate}


\subsection{Radio Environment}\label{sec:radio_environment}

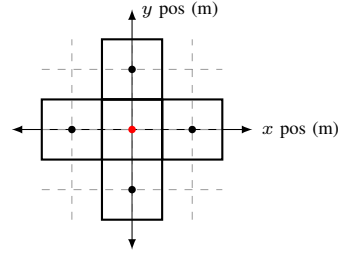
\begin{figure}[!t]
\centering
\begin{tikzpicture}[scale=0.8,font=\scriptsize]

\draw [thin, black, latex-latex] (-2,0) -- (2,0)      
        node [right, black] {$x$ pos (m)};              

    \draw [thin, black, latex-latex] (0,-2) -- (0,2)      
        node [right, black] {$y$ pos (m)};              
\draw[help lines, color=gray!80, dashed] (-1.5,-1.5) grid (1.5,1.5);
\node[circle,fill,inner sep=1pt,color=red](a) at (0,0) {};
\node[circle,fill,inner sep=1pt,color=black](x) at (-1,0) {};
\node[circle,fill,inner sep=1pt,color=black](y) at (1,0) {};
\node[circle,fill,inner sep=1pt,color=black](z) at (0,-1) {};
\node[circle,fill,inner sep=1pt,color=black](t) at (0,1) {};

  \foreach \linker / \regter in {%
    {(-0.5,-0.5)}/{(0.5,0.5)}, %
    {(-0.5,0.5)}/{(0.5,1.5)}, %
     {(-0.5,0.5)}/{(0.5,-1.5)}, %
     {(-0.5,0.5)}/{(-1.5,-0.5)}, %
    {(0.5,0.5)}/{(1.5,-0.5)}%
  } { \draw[thick] \linker rectangle \regter; }
 
\end{tikzpicture}
\caption{Radio environment.  The red point represents the serving cell.  The black points in the adjacent squares are the low power nodes.}
\label{fig:network}
\end{figure}

The radio environment is an \textit{orthogonal frequency-division multiplexing} (OFDM) indoor cellular cluster using frequency division duplex and multiple access with one tier of neighboring small cells each with a square geometry length $L$ as shown in Fig.~\ref{fig:network}.  The UEs are scattered in $\mathbb{R}^2$ according to a \textit{homogeneous poisson point process} (PPP) \cite{bacelli}. This process $\Phi$ has an \textit{intensity parameter} $\lambda$ which represents the expected number of users served by the small cell per unit area.   We define the point process $\Phi$ by the number of stationary users $N$ in the service area of the small cell $W$ sampled from a Poisson distribution with mean $\lambda W = \lambda L^2$.  The $i$-th UE coordinates are sampled from an \textit{independent and identically distributed} (i.i.d.) continuous uniform distribution.  There are $N_\text{UE}$ UEs per cell.  We make them stationary to increase our channel coherence time for reinforcement learning purposes.

We choose the square geometry because in an indoor environment with a typical omnidirectional cell installed at the center of square-shaped floor plans, a square tessellation is possible.  This is because as the transmitted signals are attenuated further more towards the square forming walls.  In practice, square geometry has been used in indoor commercial deployments \cite{cisco}, and is therefore our choice.

This cellular cluster can be in a normal state or undergo some fault-generated actions.  These faults $\mathcal{N}$ cause the channel impairment and are tracked in a special register.  We show these faults in Table~\ref{table:network_actions}.

We start by writing the signal model in an additive white Gaussian noise channel for our indoor system
\begin{equation}
y_i[t] = h_i[t] s_i[t] + n[t], \qquad i = 1, 2, \ldots\, N_\text{UE}
\label{eq:signal}
\end{equation}
where $y_i[t]$ is the received signal for the $i$-th UE, $h_i[t]$ is a single-tap channel, and $n[t]$ is a Gaussian random process sampled from $\mathrm{Norm}(0, \sigma_n^2)$.

\begin{table}[!t]
\setlength\doublerulesep{0.5pt}
\caption{Network Actions $\mathcal{N}$}
\label{table:network_actions}
\vspace*{-0.1in}
\centering
\begin{threeparttable}
\begin{tabular}{ cl|c} 
\hhline{===}
Action $\nu$ & Definition & Rate \\
\hline 
0 & Cluster is normal. & $p_0$ \\
1 & Feeder fault alarm (3 dB loss of signal). &$p_1$\\
2 & Neighboring cell down. &$p_2$\\
3  & VSWR out of range alarm.&$p_3$\\
4 & Feeder fault alarm cleared.\tnote{\textdagger} &$p_4$\\
5 & Neighboring cell up again.\tnote{\textdagger} &$p_5$\\
6  & VSWR back in range.\tnote{\textdagger} &$p_6$\\
\hline
\hhline{===}
\end{tabular}
\begin{tablenotes}\footnotesize
\item[\textdagger] These actions cannot happen if their respective alarm did not happen first.
\vspace*{-0.2in}
\end{tablenotes}
\end{threeparttable}
\end{table}

Now, we compute the received downlink SINR for the $i$-th UE at TTI $t$ ($\gamma_{\text{DL}, i}[t]$) for $i = 1, 2, \ldots, N_\text{UE}$ as:

\begin{align}\label{eq:sinr_2}
    \gamma_{\text{DL}, i} \triangleq \frac{\vert y_i\vert^2}{\sigma_n^2 + \underbrace{\sum_{j: \mathbf{o}_j \in\mathcal{C} \setminus \{{\mathbf{o}_0}\}} k_j \vert y_j\vert^2}_{\text{ICI}}}.
\end{align}
where  $\mathcal{C}$ is a set of all the cells in the cluster and $\mathbf{o}_j$ is the coordinates of the $j$-th base station. Without loss of generality, we assume that $\mathbf{o}_0$ is the serving cell placed at the origin, $k_j \ge 0$ is the proportion of users from the adjacent cells $j$ whose signals are transmitted on the same PRB as the $i$-th UE at TTI $t$.  Those signals therefore cause \textit{inter-cell interference} (ICI).

The forward link budget at any TTI $t$ is written as:
\begin{equation}
P_{\text{UE},i}[t] =  P_\text{TX}[t] + G_\text{TX} - L_\text{m} - L_{\text{a},i}[t]  + G_\text{UE}
\label{eq:linkbudget}
\end{equation}
where $P_{\text{UE},i}$ is the received power for the allocated \textit{physical resource blocks} (PRB) transmitted at power $P_\text{TX}$, $G_\text{TX}$ is the antenna gain of the transmitter, $L_\text{m}$ is a miscellaneous loss (e.g., feeder loss and return loss), $L_{\text{a},i}$ is the path loss over the air interface for line of sight indoor propagation for the $i$-th user, and $G_\text{UE}$ is the UE antenna gain.

\begin{align}\label{eq:sinr_final}
    \gamma_{\text{DL}, i}[t] \triangleq \frac{P_{\text{UE},i}[t] }{\sigma_n^2 + \sum_{j: \mathbf{o}_j \in\mathcal{C} \setminus \{{\mathbf{o}_0}\}}  k_jP_{\text{UE}, j}[t] }.
\end{align}

The effective downlink received SINR for users in the serving cell $\mathbf{o}_0$ at a given TTI $t$, $\bar{\gamma}_{\text{DL}}[t]$ in dB is
\begin{equation}
\label{eq:eff_sinr}
\bar{\gamma}_{\text{DL}}[t] \triangleq 10\log \left ( \frac{1}{N_\text{UE}} \sum_{i = 1}^{N_\text{UE}} \gamma_{\text{DL}, i}[t] \right )  \qquad\text{(dB)}
\end{equation}
which is the quantity to improve (i.e., our \textit{objective}).

\subsection{Reinforcement Learning}
\label{sec:ql}

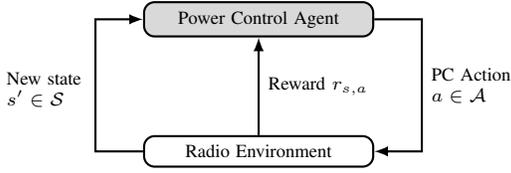
\begin{figure}[!t]
\centering
\begin{adjustwidth}{1cm}{0cm}
\begin{tikzpicture}[node distance = 5em, auto, thick, scale=0.9, font=\scriptsize]]
    \node [rectangle, draw, 
    text width=8em, text centered, rounded corners, minimum height=1em,fill=gray!30] (Agent) {Power Control Agent};
    \node [rectangle, draw, 
    text width=8em, text centered, rounded corners, minimum height=1em, below of=Agent] (Environment) {Radio Environment};
    
     \path [draw, -latex] (Agent.0) --++ (2em,0em) |- node [text width=5em,near start]{PC Action\\ $a\in\mathcal{A}$} (Environment.0);
     \path [draw, -latex] (Environment.180) --++ (-2em,0em) |- node [text width=5em, xshift=-0.3cm,yshift=-1.3cm] {New state\\ $s^\prime\in\mathcal{S}$} (Agent.180);
     \path [draw, -latex] (Environment.north) --++ (0em,0em) -- node [right] {Reward $r_{s,a}$} (Agent.south);
\end{tikzpicture}
\vspace*{-0.2in}
\caption{Reinforcement learning elements.}
\label{fig:reinf}
\end{adjustwidth}
\end{figure}

We formulate the VoLTE closed loop PC problem as a reinforcement learning based problem using Tabular $Q$-learning as shown in Algorithm~\ref{alg:the_pc_alg}.  The objective is to meet the target effective SINR $\bar{\gamma}_\text{DL, target}$ despite signal impairments, which are tracked in a register $\bm{\varphi}_\text{fault}[t]$ where $\bm{\varphi}_\text{fault}\in\mathbb{F}_2^{\vert\mathcal{N}\vert}$.  The set of actions carried out by the agent is $\mathcal{A} = \{a_i\}_{i = 0}^{n-1}$ and the set of states is $\mathcal{S} = \{s_i\}_{i = 0}^{m-1}$.   The environment grants the agent a reward $r_{s,a}$ after the algorithm takes an action $a\in\mathcal{A}$ when it is in state $s\in\mathcal{S}$ at discrete time $t$.  These elements are shown in Fig.~\ref{fig:reinf}.

Following \cite{Sutton}, we denote $Q(s, a)$ as the state-action value function, i.e., the expected discounted reward when starting in state $s$ and selecting an action $a$. To derive $Q(s,a)$, we build an $m$-by-$n$ table $\mathbf{Q}\in\mathbb{R}^{m\times n}$.  This allows us to use the shorthand notation: $Q(s,a) \triangleq [\mathbf{Q}]_{s,a}$ and write:
\begin{equation}
    \label{eq:bellman_tabular}
    Q(s,a) \triangleq (1-\alpha) Q(s,a) +  \alpha \left [ r_{s,a} + \gamma \max_{a^\prime} Q(s^\prime,a^\prime) \right ]
\end{equation}
where $\alpha: 0 < \alpha < 1$ is the learning rate and $\gamma: 0 < \gamma < 1$ is the \textit{discount factor} and determines the importance of the predicted future rewards.  The next state is $s^\prime$ and the next action is $a^\prime$.
For our proposed closed loop PC algorithm, the the upper bound of the time complexity for $Q$-learning is in $\mathcal{O}(m^2)$ \cite{Koenig1993}.

An \textit{episode} is a period of time in which an interaction between the agent and the environment takes place. In our case, this period of time is $\tau$ TTIs. During an episode $z: z\in \{0,1, \ldots, \zeta\} $, the agent makes the decision to maximize the effects of actions decided by the agent. To do so, we use the $\epsilon$-greedy strategy for learning, where $\epsilon$ is the \textit{exploration rate} and serves to select a random action $a\in\mathcal{A}$ with a probability $\epsilon: 0 < \epsilon < 1 $, known as exploration, as opposed to selecting an action based on previous experience, which is also known as exploitation. The exploration rate decays in every episode until it reaches $\epsilon_\text{min}$.

We list the actions performed by the PC agent (known as \textit{PC commands}) in Table~\ref{table:actions}.   The PC agent attempts to reach $\bar{\gamma}_\text{DL, target}$ through a series of PC commands $c$ in reaction to various impairments due to the network actions $\nu\in\mathcal{N}$ during any given TTI $t$.   These actions have a finite impact on the effective downlink received SINR $\bar\gamma_\text{DL}[t]$.

After each action, we compute the effective downlink SINR gain (or loss), which is $\Delta\bar\gamma_\text{DL}[t]$, as
\begin{equation}
\label{eq:sinr_diff}
\Delta\bar\gamma_\text{DL}[t] \triangleq \bar\gamma_\text{DL}[t] - \bar\gamma_\text{DL}[t - 1].
\end{equation}

\begin{proof}[Computation of contribution of actions $\nu=1, 3$]
When the \textit{voltage standing wave ratio}  (VSWR) changes from $v_0$ to $v$ in TTI $t$ (action $\nu=3$), we compute the change in signal loss due to return loss, which is equal to the change in SINR, as
\begin{equation}
\Delta L  = 10\log \left (\bigg\vert\frac{v_0 + 1}{v_0 - 1}\bigg\vert\cdot \bigg\vert\frac{v - 1}{v + 1}\bigg\vert\right ) ^2.
\end{equation}
Action $\nu = 1$ is a special case with $\Delta L = -3\,\text{dB}$.
\end{proof}

\begin{proof}[Computation of contribution of action $\nu = 2$]
When the neighbor cell $k$ is down, the transmit power of the adjacent cells $j$ will increase by an arbitrary quantity $0 < \varepsilon_j \le 1$ and the number of interferers will decrease.  Therefore, for simplicity of computation, we derive the lower bound of the downlink effective SINR of this action $\nu=2$ as
\begin{equation}
\begin{aligned}
\bar\gamma_\text{DL} &= 
\frac{P_{\text{UE},i}}{\sigma_n^2 + \sum_{j: j \in\mathcal{C} \setminus \{ \mathbf{o}_0, \mathbf{o}_\ell\}} (1+\varepsilon_j) k_j P_{\text{UE}, j} } \\
& \stackrel{(a)}{\ge} \frac{ P_{\text{UE},i}}{\sigma_n^2 + \vert{\mathcal{C} \setminus  \{ \mathbf{o}_0, \mathbf{o}_\ell\}\vert }  P_{\text{BS}}^{\rm max}  } \\
& \stackrel{(b)}{=}\frac{ P_{\text{UE},i}}{\sigma_n^2 + (\vert\mathcal{C}\vert - 2)  P_{\text{BS}}^{\rm max}  }
\end{aligned}
\end{equation}
where  $P_\text{BS}^{\rm max}$ is the maximum transmit power of the indoor cell.
$(a)$ is since we use the maximum small cell transmit powers instead of the increased received power measured at the UE, and $(b)$ is since the cardinality of $\mathcal{C}$ is reduced by two: the serving and neighbor cells from step $(a)$.
\end{proof}

\begin{proof}[Computation of contribution of actions $\nu=4,5,6$]
These actions are a result of their respective fault actions being cleared.  Therefore, we reverse the effect of actions $1, 2,$ and $3$ respectively.
\end{proof}

\begin{table}[!t]
\setlength\doublerulesep{0.5pt}
\caption{Power Control (PC) Algorithm Actions at time $t$}
\vspace*{-1em}
\label{table:actions}
\centering
\begin{tabular}{ cl } 
\hhline{==}
Action $a$ & Definition \\
\hline 
0 & $c=0$. \\
1 & $c=-1$  repeated three times (i.e., $\eta[t] = 3$). \\
2 & $c=-1$  repeated one time (i.e., $\eta[t] = 1$). \\
3 &  $c=+1$ repeated one time. \\
4 & $c=+1$ repeated three times.\\
\hhline{==}
\end{tabular}
\end{table}

\begin{table}[!t]
\setlength\doublerulesep{0.5pt}
\caption{Power Control (PC) Algorithm States}
\vspace*{-1em}
\label{table:states}
\centering
\begin{tabular}{ cl } 
\hhline{==}
State $s$ & Definition \\
\hline
0 & Transmit power is unchanged (i.e., $c=0$). \\
1 & Transmit power is increased (i.e., $c= +1$). \\
2 & Transmit power is decreased (i.e., $c= -1$). \\
\hhline{==}
\end{tabular}
\end{table}

 \begin{algorithm}[!t]
        \small
        \caption{\small \mbox{VoLTE Downlink Closed Loop Power Control}}
        \label{alg:the_pc_alg}
        \DontPrintSemicolon
        \KwIn{Initially computed effective downlink SINR value ($\bar{\gamma}_{\text{DL},0}$) and desired target effective SINR value ($\bar\gamma_\text{DL, target}$).}
        \KwOut{Optimal sequence of power commands required to achieve the target SINR value during a VoLTE frame $z$, which has a duration of $\tau$ amid network impairments captured in $\bm{\varphi}_\text{fault}$.}
        Define the power control (PC) actions $\mathcal{A}$, the set of PC states $\mathcal{S}$, the exploration rate $\epsilon$, the decay rate $d$, and $\epsilon_\text{min}$.\; 
        $t := 0$\; 
        $\bar{\gamma}_\text{DL} := \bar{\gamma}_\text{DL,0}$\;
        $(s, a) := (0,0)$ \;
        $\bm{\varphi}_\text{fault} := [0,0,\ldots,0]$\;
        \Repeat  {$\bar{\gamma}_\text{DL} \ge$ $\bar\gamma_\text{DL, target}\;\mathrm{or}\; t \ge\tau$} {
           $t := t  + 1$\;
           $\epsilon := \max(\epsilon\cdot d, \epsilon_\text{min})$\;
           Sample $r \sim \mathrm{Uniform}(0,1)$\;
          \eIf {$r \le \epsilon$} {
           Select an action $a \in \mathcal{A}$ at random.\;
          } {
           Select an action $a \in \mathcal{A}, a = \arg\max_{a^\prime} Q(s,a^\prime)$.\;
          }
           Perform action $a$ (power control) on $P_\text{TX}[t]$ and obtain reward $r_{s,a}$.\;
           Observe next state $s^\prime$.\;
           Update the table entry $Q(s,a)$ as in (\ref{eq:bellman_tabular}). \;
           $s := s^\prime$ \;
        }
        Proceed to the next VoLTE frame $z+1$.\;
        \end{algorithm}

\section{Power Control Algorithms} \label{sec:power_control}
\subsection{Fixed Power Allocation} \label{sec:fixed_pa}

The \textit{fixed power allocation} (FPA) power control method is an open-loop PC algorithm that serves as a baseline for comparison.   It is a common power allocation scheme where the total transmit power is simply divided equally among all PRBs in the operating band  $N_\text{PRB}$ and is therefore constant but cannot exceed the maximum transmission power of the small cell:
\begin{equation}
P_\text{TX}[t] \triangleq P_\text{BS}^\mathrm{max} - 10\log N_\text{PRB}\qquad\text{(dBm)}  \\
\end{equation}

\subsection{Closed Loop Power Control} \label{sec:ql_pc}
Owed to the closed loop PC, we can write $P_\text{TX}$ in dBm at any given TTI $t$ as: 
\begin{equation}
\nonumber
\begin{aligned}
P_\text{TX} [t] = \min\! \big (P_\text{BS}^{\rm max},P_\text{TX}[t - 1] + \eta[t]c[t]\big) \qquad\text{(dBm)}
\end{aligned}
\end{equation}
where   $\eta[t]$ is the repetition factor of a power command $c$ in a given TTI $t$.  Power control cannot cause the transmit power to exceed the maximum transmit power of the serving cell.  Power commands can be issued multiple times per TTI as shown in Table~\ref{table:actions}.  These power commands impact the algorithm state as shown in Table~\ref{table:states}.

\section{Voice Call Performance Metrics}\label{sec:benchmark}
We use two performance metrics: \textit{call retainability} and \textit{mean opinion score} (MOS) to compare both algorithms.

\subsection{Call Retainability}
We define call retainability for the radio environment as a function of an effective downlink SINR threshold $\bar{\gamma}_\text{DL, min}$ obtained during the final episode $\zeta$:
\begin{equation}
\text{Retainability}  \triangleq 1 - \frac{1}{\tau}\sum_{t = 0}^{\tau} \mathbbm{1}_{\bar\gamma[t] \le \bar{\gamma}_\text{DL, min}} .
\label{eq:retainability}
\end{equation}
\subsection{Mean-Opinion Score}

To benchmark the audio quality, we compute \textit{mean-opinion score} (MOS) using the experimental MOS formula \cite{Yamamoto97impactof}.  We obtain the packet error rate from the simulation over $\tau$ frames in the final episode $\zeta$ using the symbol probability of error of a QPSK modulation in OFDM.  We choose a VoLTE data rate of 23.85 kbps and a voice \textit{activity factor} (AF), the ratio of voice payload to silence during a voice frame, of 0.7. We refer to the source code \cite{mycode} for further details.  Result is in Fig.~\ref{fig:mos}.

\vspace*{0.1in}
\section{Simulation Results}\label{sec:results}

\begin{table*}[!t]
\setlength\doublerulesep{0.5pt}
\caption{Radio Environment Parameters}
\vspace*{-1em}
\label{table:rf_parameters}
\centering
\begin{tabular}{ lrlr } 
\hhline{====}
Parameter & Value & Parameter & Value \\
 \hline
LTE bandwidth & 20 MHz & Base station maximum power  $P^\textrm{max}_\text{LPN}$ & 33 dBm \\ 
Downlink center frequency & 2.6 GHz& Base station initial power setting & 13 dBm \\ 
Maximum number of UEs per serving cell $N_\text{UE}$ &  10 & Antenna model & omnidirectional \\
Number of physical resource blocks $N_\text{PRB}$ & 100 & Antenna gain $G_\text{TX}$ & 16 dBi \\ 
Cellular geometry & square ($L$ = 10 m) &Antenna height  & 10 m\\ 
Propagation model & COST 231 & User Equipment (UE) antenna gain & -1 dBi \\ 
Propagation environment & indoor  & UE height  & 1.5 m \\

\hhline{====}
\end{tabular}
\end{table*}

\begin{table}[!t]

\caption{Machine Learning Parameters}
\label{table:ml_parameters}
\setlength\doublerulesep{0.5pt}
\vspace*{-0.1in}
\centering
\begin{tabular}{ lr } 
\hhline{==}
Parameter & Value \\
\hline
Number of episodes $\zeta$ & 707 \\
One episode duration $\tau$ (ms) & 20 \\
Discount factor $\gamma$ & 0.950 \\
Exploration rate $\epsilon$ & 1.000\\
Minimum exploration rate $\epsilon_\text{min}$ & 0.010\\
Exploration rate decay $d$ &  0.99 \\
Learning rate  $\alpha$ & 0.001 \\
Number of states  & 3 \\
Number of actions   & 5 \\
\hhline{==}
\end{tabular}
\end{table}


We implement Algorithm \ref{alg:the_pc_alg} and set the machine learning parameters as in Table~\ref{table:ml_parameters}.  To examine the worst case scenario, we set the probabilities in Table~\ref{table:network_actions} as: $p_0 = 5/11, p_1 = p_2 = p_3 = \ldots = p_6 = 1/11$.  This way we give all faults an equally likely chance of occurrence while having the network perform reliably at least for 45\% of the time.  We further set the rewards as:
\begin{equation}
r_{s,a}[t] \triangleq \begin{cases} 
r_\text{min}, & \; \bar\gamma_\text{DL}[t] = \bar\gamma_\text{DL, target}  \, \text{not feasible or}\, t\ll\tau \\
	-1, & \; \bar\gamma_\text{DL}[t] < \bar\gamma_\text{DL}[t-1]   \\
	    0, & \;  \bar\gamma_\text{DL}[t] = \bar\gamma_\text{DL}[t-1]   \\
      1, &\;  \bar\gamma_\text{DL}[t] > \bar\gamma_\text{DL}[t-1]  \\
r_\text{max}, & \;  \bar\gamma_\text{DL}[t] = \bar\gamma_\text{DL, target}  \, \text{is met.}
   \end{cases}
   \label{eq:rewards_volte_sim}
\end{equation}

For the retainability, we choose $\bar{\gamma}_\text{DL, min} = 0\,\text{dB}$ in (\ref{eq:retainability}). The radio network parameters are set as in Table~\ref{table:rf_parameters}.  We set $\bar\gamma_{\text{DL},0}$ to $4$ dB and $\bar\gamma_\text{DL, target}$ to $6$ dB.

\begin{figure}[!t]
\centering
\includegraphics[width=2.5in]{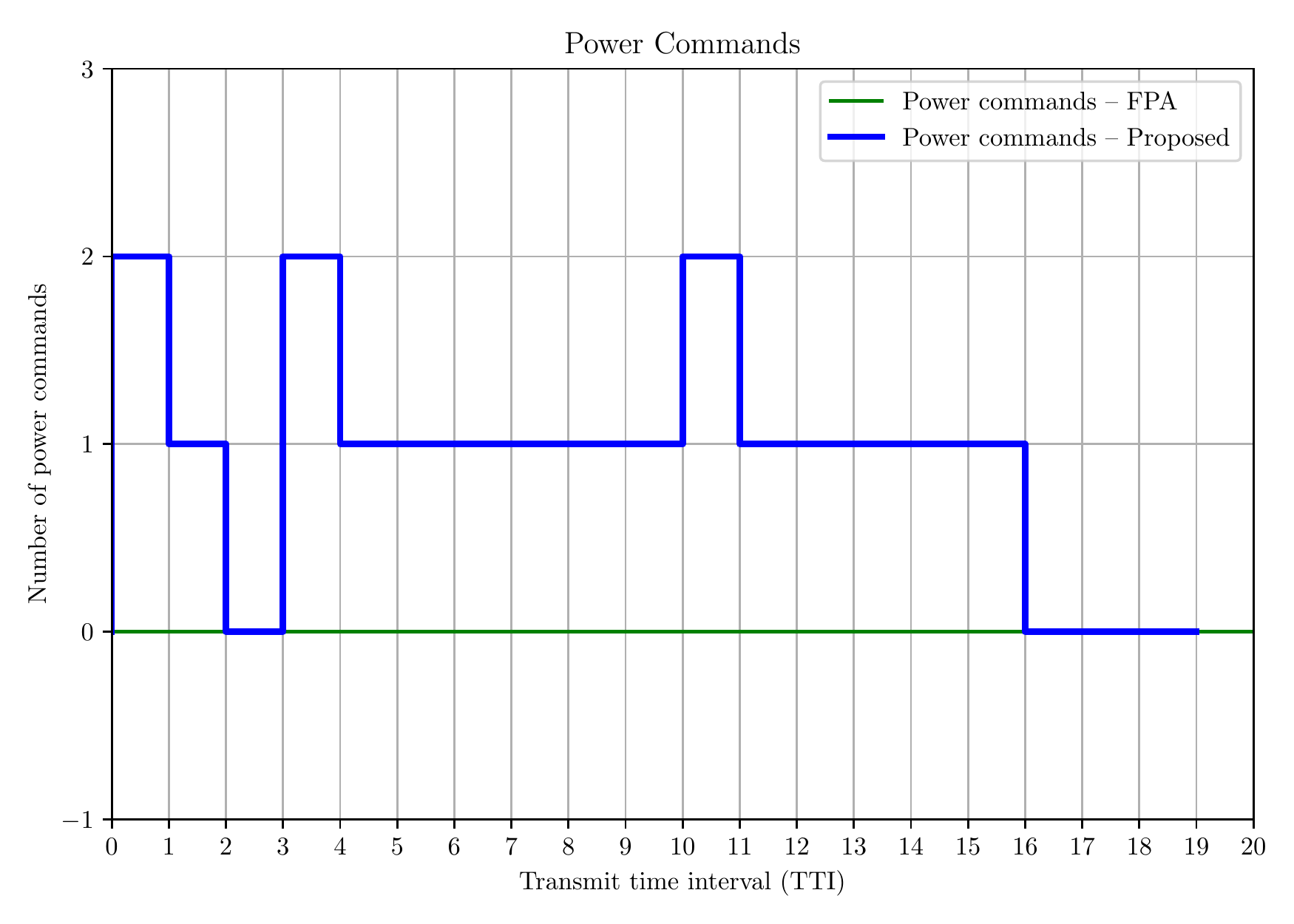}
\vspace*{-1em}
\caption{Power control (PC) sequence during the final episode $\zeta$.  Unlike fixed power allocation (FPA), our proposed closed loop power control sent several PCs per transmit time interval (TTI) for the entire VoLTE frame.}
\label{fig:pc}
\end{figure}
\begin{figure}[!t]
\centering 
\includegraphics[scale=0.48]{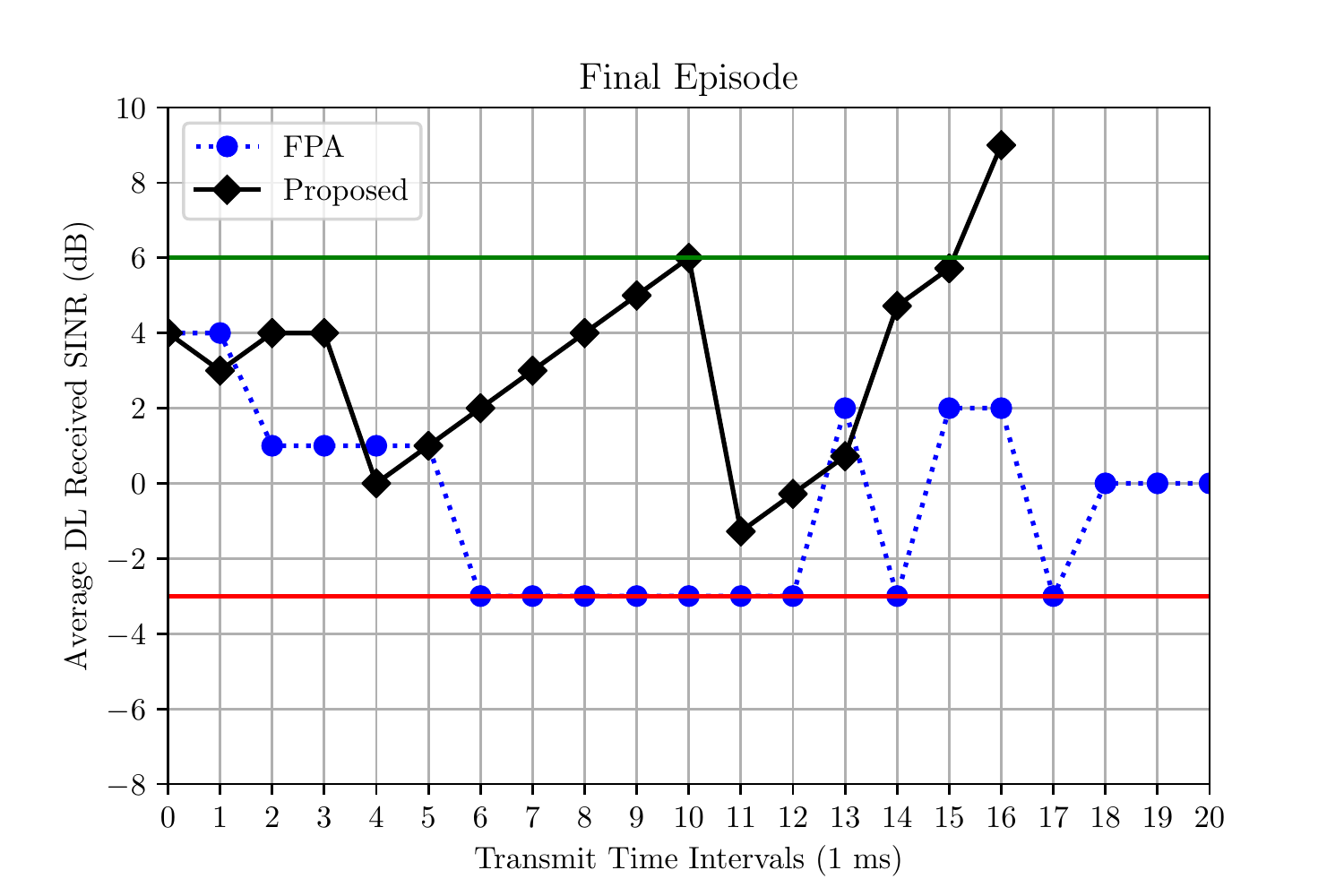}
\vspace*{-0.1in}
\caption{Downlink signal to interference plus noise ratio (DL SINR) improvement vs. simulation time for both our proposed closed loop power control using $Q$-learning and fixed power allocation (FPA) for the final episode $\zeta$. Green and red lines are  $\bar\gamma_\text{DL, target}$ and  $\bar\gamma_\text{DL, min}$ respectively. The proposed algorithm reaches the target while FPA does not.}
\label{fig:episode826}
\end{figure}
\begin{figure}[!t]
\centering
\includegraphics[width=0.45\textwidth,height=5.5cm]{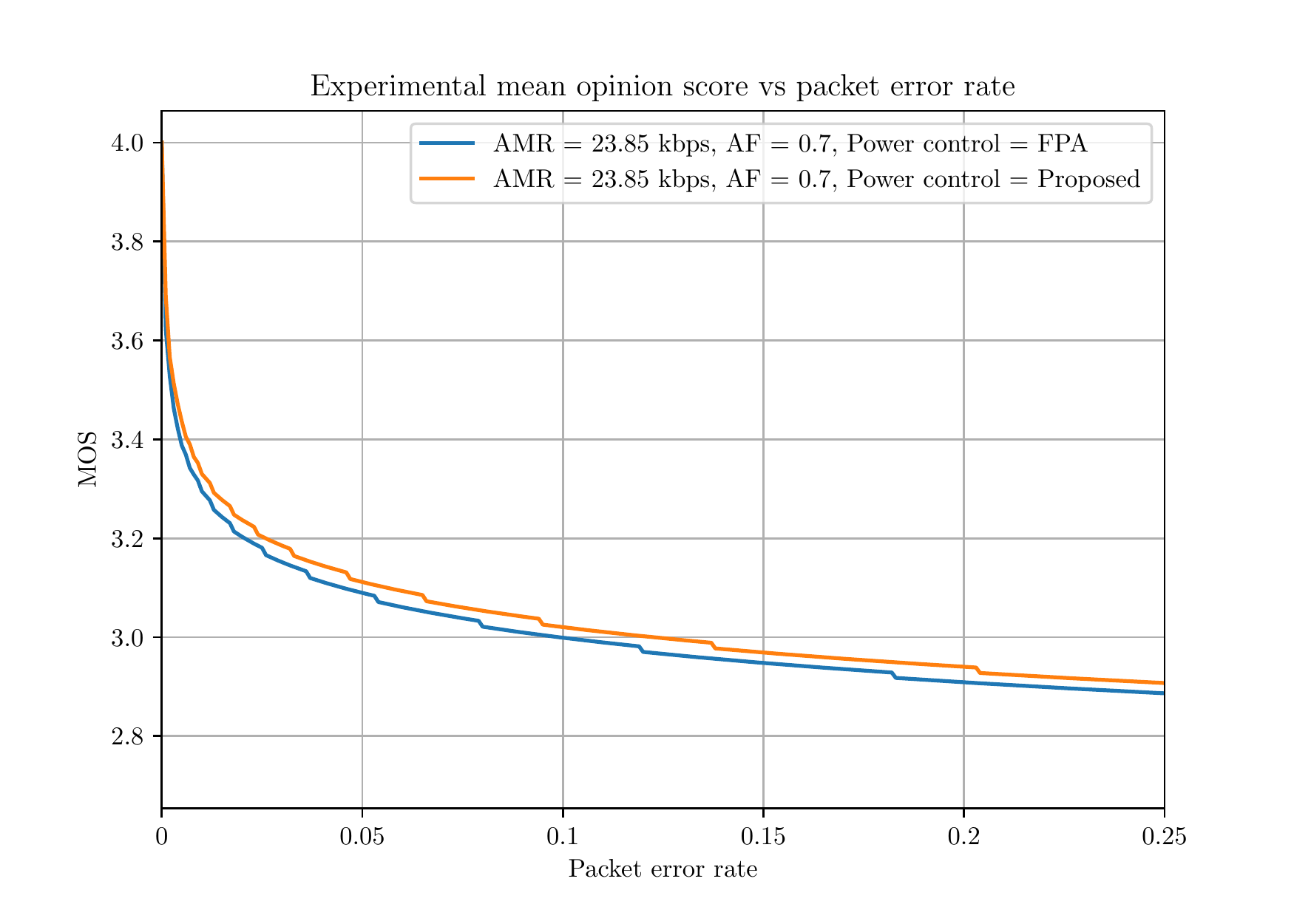}
\vspace*{-0.2in}
\caption{Mean opinion score (MOS) based on the  voice packet error rate and the experimental formula \cite{Yamamoto97impactof}.  Our proposed closed loop power control algorithm has improved MOS compared to Fixed Power Allocation (FPA).}
\label{fig:mos}
\end{figure}

The optimal $Q$\nobreakdash-learning action-value function (\ref{eq:bellman_tabular}) is learned after $\zeta$ episodes.  At this stage, the closed loop PC performs better than FPA.
Fig.~\ref{fig:pc} shows the power command sequence for the final episode $z = \zeta = 707$, where the closed loop PC algorithm causes the base station to change its transmit power to meet the desired DL SINR target, unlike FPA where no power commands are sent.  We show both algorithms in Fig.~\ref{fig:episode826} for the final episode $\zeta$.  The closed loop PC pushes the effective DL SINR to the target through an optimal sequence of power commands.  The improved retainability and experimental MOS scores due to the closed loop power control algorithm are shown in Table~\ref{table:retainability} and Fig.~\ref{fig:mos} respectively.  The reason why retainability and MOS have improved is understood directly from the impact of the effective DL SINR which increases the quantity of (\ref{eq:retainability}) and decreases the packet error rate---the main component in the experimental MOS formula.    We refer to the source code \cite{mycode} for further implementation details.

\begin{table}[!t]
\setlength\doublerulesep{0.5pt}
\caption{Retainability}
\label{table:retainability}
\vspace*{-.1in}
\centering
\begin{tabular}{ ccc} 
\hhline{===} 
 & Fixed Power Allocation & Proposed \\
\hline
Retainability & 55.00\% & \textbf{78.75\%} \\ 
\hhline{===}
\end{tabular}
\end{table}

\section{Conclusion}\label{sec:conclusions}
We introduced downlink closed loop power control using $Q$\nobreakdash-learning, which improved VoLTE performance in a realistic indoor environment compared to the open-loop fixed power allocation power control. It resulted in improvement in the quality of experience measured by the voice call retainability and MOS metrics.   This was due to the robustness of maintaining the target DL SINR.  The ability to maintain this target helps prevent a voice call from dropping and reduces the voice packet errors. 



%


\ifCLASSOPTIONcaptionsoff
  \newpage
\fi

\IEEEtriggeratref{6}


%

\bibliography{references}  

\bibliographystyle{IEEEtran}

\end{document}